\documentclass[amsmath,amssymb,twocolumn,showpacs,superscriptaddress,prl]{revtex4}

\usepackage{amsmath,amsthm,amssymb,bm}

\newcommand{\be}{\begin{equation}}
\newcommand{\ee}{\end{equation}}

\newcommand{\fa}{\Psi}

\newcommand{\fad}{\Psi^\dagger}

\newcommand{\lam}{\lambda}
\newcommand{\ra}{\rangle}
\newcommand{\la}{\langle}
\newcommand{\inti}{\int_{-\infty}^{+\infty}}

\newcommand{\Imm}{\Im }
\newcommand{\Ree}{\Re }

\begin{document}

\title{Large-Distance Asymptotic Behavior  of the Correlation Functions of 1D
Impenetrable Anyons at Finite Temperatures}

\author{Ovidiu I. P\^{a}\c{t}u}

\affiliation{C.N. Yang Institute for Theoretical Physics, State
University of New York at Stony Brook, Stony Brook, NY 11794-3840,
USA }
\affiliation{Institute for Space Sciences, Bucharest-M\u{a}gurele, R
077125, Romania}

\author{Vladimir E. Korepin}

\affiliation{C.N. Yang Institute  for Theoretical Physics, State
University of New York at Stony Brook, Stony Brook, NY 11794-3840,
USA }

\author{Dmitri V. Averin}
\affiliation{Department of Physics and Astronomy, State University
of New York at Stony Brook, Stony Brook, NY 11794-3800, USA }
\email[Electronic addresses: ]{ipatu@grad.physics.sunysb.edu ;
korepin@max2.physics.sunysb.edu ; dmitri.averin@stonybrook.edu}

\begin{abstract}

The large-distance asymptotic behavior  of the field-field correlators has
been computed for one-dimensional impenetrable anyons at finite
temperatures. The asymptotic behavior agrees with the predictions of
 conformal field theory at low temperatures and reproduces
the known results for impenetrable bosons and free fermions in
appropriate limits. We have also obtained an integrable system of
partial nonlinear differential equations which completely
characterizes the 2-point correlation functions. The system is the
same as for bosons but with different initial conditions.

\end{abstract}


\maketitle


Anyonic exchange statistics is defined naturally for impenetrable
hard-core particles in the two-dimensional space \cite{LM,FW}, which
is sufficiently ``large'' to allow particles to rotate around each
other, but, on the other hand, is ``small'' enough for the particle
trajectories to be entangled by such rotations, so that the winding
number of the rotations is a well-defined characteristics of each
trajectory. By contrast, one-dimensional (1D) space seems too small
to support any exchange statistics, since  impenetrable particles
can not be exchanged at all when confined to move on a line. This
simple expectation, frequently expressed in the literature, is not
fully correct because of two circumstances. One is that in
the case of quasi-periodic boundary conditions (i.e., motion on a
circle), the particles can be exchanged indirectly by transfer along
the circle. These indirect exchanges mean that the statistics
affects the precise form of the boundary conditions and, therefore,
the position of the quasiparticle energy levels, making the
thermodynamic properties of the system of $N$ impenetrable particles
statistics-dependent at the level of $1/N$ corrections
\cite{ZW,BGO,MDG,PKA}. Even more importantly, in all realistic 1D
systems, particles can be added to or removed from the system,
processes which are described by the field-field correlation
functions, and are strongly sensitive to the statistics even in the
limit $N\rightarrow \infty$. Indeed, due to fluctuations in  the 1D
quantum liquid, its multi-particle wavefunctions contain terms with
different distribution of particles relative to a fixed point $x$,
where the external particle is added. This means that the amplitudes
of the addition/removal processes, and associated field-field
correlators unavoidably involve contributions with different
ordering of particles, and therefore depend on the exchange
statistics \cite{AN}. The purpose of this work is to present exact
calculation of the long-distance asymptotics of the field-field
correlators for the system of 1D impenetrable hard-core anyons at
arbitrary finite temperatures,  $0<T<\infty$. In practice, the
anyonic liquid studied in this work is similar to the systems that
can be realized with edges of the electron liquids in the FQHE
regime.

We consider the most basic model of 1D anyons \cite{AK}, the
Lieb-Liniger gas, characterized by the second-quantized Hamiltonian
\begin{eqnarray}
H&=&\int dx \left [\partial_x \fad(x)\partial_x\fa(x)
+c\fad(x)\fad(x)\fa(x)\fa(x) \right. \nonumber \\ & &\ \
\;\;\;\;\;\;\; \left. -h\, \fad(x)\fa(x) \right] , \label{hama}
\end{eqnarray}
where $c>0$ is the coupling constant and $h$ is the chemical
potential. We limit ourselves to the case of impenetrable particles,
$c \rightarrow \infty$. Anyonic exchange statistics is introduced by
imposing the following commutation relations on the field operators:
\begin{eqnarray}
\fa(x_1)\fad(x_2)&=&e^{-i\pi\kappa\epsilon(x_1-x_2)} \fad(x_2)
\fa(x_1)+ \delta(x_1-x_2),\nonumber\\
\fa(x_1)\fa(x_2)&=&e^{i\pi\kappa\epsilon(x_1-x_2)} \fa(x_2)
\fa(x_1)\, , \label{com1} \end{eqnarray}
where $\kappa \in [0,1]$ is the statistics parameter, and
$\epsilon(x)=x/|x|,\, \epsilon(0)=0.$ The commutation relations
(\ref{com1}) become bosonic for $\kappa=0$, and fermionic for
$\kappa=1$. They imply that the multi-particle wavefunction acquires
the phase factor $e^{\pm i\pi \kappa}$ upon the exchange of any of
the two nearest-neighbor particles, where the signs $\pm$ need to be
assigned consistently for all pairs of the wavefunction coordinates
\cite{AN,PKA}. Thermodynamic properties of the Lieb-Liniger gas of
anyons were studied in \cite{ZW,BGO,MDG,PKA,BGH}, and for $c
\rightarrow \infty$, coincide (neglecting $1/N$ terms) with those of
the free fermions. At $T=0$, the quasiparticle momenta fill out the
Fermi sea with the Fermi momentum and velocity given by,
respectively, $k_F=\sqrt{h}$ and $v_F=2\sqrt{h}$, in conventions of
Eq.~(\ref{hama}). Here, we are interested in the equal-time
field-field correlator defined as usual by
\be \la \fad(x_1)\fa(x_2)\ra \equiv \frac{\mbox{Tr} \left( e^{-H/T}
\fad(x_1) \fa(x_2) \right)}{\mbox{Tr} e^{-H/T} }\, . \label{corr}
\ee
The correlator depends only on the difference $x \equiv (x_1-x_2)$ of
the coordinates, and since $\la \fad(x_2)\fa(x_1)\ra =\la\fad(x_1)
\fa(x_2) \ra^*$, we can consider only the range $x>0$.

As the reference point for the results of this work, we mention that
at $T\rightarrow 0$, the long-distance asymptotics of the correlator
(\ref{corr}) can be obtained from the standard bosonization approach
\cite{FDMH}. The particle operators are expressed in it
through two bosonic fields: the integral $\theta(x)$ of the
long-wave part of the density fluctuations $\rho (x)$, i.e.,
$\partial \theta(x)/\partial x=\rho (x)/\pi $, and the conjugate
field $\phi (x)$ defined by the commutation relation $[\phi (x),
\theta (x')]=i\pi \epsilon(x-x')/2$. The usual bosonization
ideology, i.e., the operator $e^{i\phi}$ reducing density by the
amount that corresponds to one particle, and $\theta$ producing the
appropriate phase changes of the wavefunction across each particle,
implies immediately that the anyonic field operators (\ref{com1})
can be written as \cite{CM}
\be \fa (x,t) \sim \sum_{m}e^{-i(\kappa+2m) [k_Fx+\theta(x,t) ]} \,
e^{i \phi(x,t) } \, . \label{bos1} \ee
The sign $\sim$ in Eq.~(\ref{bos1}) is the reminder that the
relative amplitudes of different components are not defined in this
sum. The commutation relations between $\theta$ and $\phi$ mean
that the individual terms in (\ref{bos1}) satisfy the necessary
exchange relations (\ref{com1}) away from $x_1=x_2$.
Using Eq.~(\ref{bos1}) for the field operators, and taking the
standard average over the equilibrium fluctuations of $\theta$ and
$\phi$, one finds the long-distance behavior of the correlator
(\ref{corr}):
\begin{eqnarray}
\la \fad(x)\fa(0)\ra = \frac{b_0 \, e^{i\kappa k_Fx}}{ [\sinh(\pi T
x/v_F)]^{(1+\kappa^2)/2}} \nonumber \\  +\, \frac{b_{-1} \,
e^{i(\kappa-2) k_Fx}}{[\sinh(\pi T x/v_F)]^{[1+(\kappa-2 )^2]/2} }
\, , \label{bos2} \end{eqnarray}
where the constants $b$ are some undetermined amplitudes. Equation
(\ref{bos2}) gives the two leading terms which correspond to
$m=0,-1$ in (\ref{bos1}) and are relevant for the discussion below.
In general, the correlator contains the higher-order terms which
include the full expansion (\ref{bos1}) \cite{CM,PKA} and goes
beyond it, if one uses the somewhat more general conformal field-theory
approach \cite{PKA}.

The main new result obtained in this work is the expression for the
asymptotics of the anyonic field-field correlator (\ref{corr}) for
large distances, $x \rightarrow +\infty$, that extends the
bosonization result to arbitrary temperatures. The bosonization
(\ref{bos1}) is known to rely effectively on approximation of the
linear quasiparticle spectrum: $\epsilon_{k} = \pm v_F(k \mp k_F)$
for quasiparticle momenta $k \simeq \pm k_F$, which implies the
exact electron-hole symmetry. This means that from the physics
perspective, such an extension is needed for all phenomena (e.g.,
thermo-electric effects) which depend on the asymmetry between
electrons and holes. On the other hand, our result is also
interesting as an example of exact calculation of a non-trivial
quantity in statistical mechanics. It is:
\be \la \fad(x)\fa(0)\ra
=e^{-x\sqrt{T}C(\beta,\kappa)/2}\sum_{j=0,-1} c_j e^{ix
\sqrt{T}\lam_j} , \label{asympt} \ee
where $c_0,c_{-1}$ are some complex constants, $\beta \equiv h/T$,
and $C(\beta,\kappa)$ is defined as
\be C(\beta,\kappa)=\frac{1}{\pi}\inti\ln\Big( \frac{e^{ \lam^2
-\beta}+1}{e^{\lam^2-\beta}-e^{i\pi\kappa}}\Big)d\lam\, . \label{C}
\ee
The branch of the complex logarithm in Eq.~(\ref{C}) is fixed by the
conditions that there are no cuts intersecting the real axis of
$\lam$, and that the logarithm tends to zero at $\lam \rightarrow
\infty$. The complex constants $\lam_0$ and $\lam_{-1}$ in
(\ref{asympt}) depend on $\beta$ and $\kappa$ as $\lam_j=(-1)^j
[\beta +i\pi(\kappa+2j)]^{1/2}$, or, separating explicitly real and
imaginary parts,
\be
\lam_j=[(-1)^j(\alpha_j+\beta)^{1/2}+i(\alpha_j-\beta)^{1/2}]/\sqrt{2}
\, , \label{lamp} \ee
where $\alpha_j \equiv [\beta^2+\pi^2(\kappa+2j)^2]^{1/2}$. Strictly
speaking, the accuracy of our calculation allows us to keep the
second term in (\ref{asympt}) only when it does not decay much
faster than the first: $\Imm \lam_{-1}<2\Imm \lam_0$,   condition
satisfied when the statistics parameter $\kappa$ is not too small.
For $\kappa$ approaching the fermionic point $\kappa=1$, the two
terms have the same decay rate $\Imm \lam_0\sim \Imm\lam_{-1}$, and
$\Ree \lam_0\sim -\Ree \lam_{-1}$ -- see Eq.~(\ref{lamp}), so that
the asymptotics exhibits beats with an exponentially decaying
envelope.


The method we used to obtain Eq.~(\ref{asympt}) (described below)
allows us to find both the exponents and the amplitudes $c_j$ in
this equation. More technical results for the amplitudes will be
presented in the follow-up publication. Here we focus on the main
exponential factors. We start the discussion with the {\em analysis
of $C(\beta,\kappa)$ } (\ref{C}) in different limits. In the ``gas
phase'', when the chemical potential $h$ of the anyons is negative,
i.e. $\beta = h/T <0$, we can split the logarithm in Eq.~(\ref{C})
in two and use for them the expansion $\ln(1+z)=\sum_{n=1}^\infty
(-1)^{n+1}z^n/n$ at $|z|<1$, to obtain:
\be C(\beta,\kappa) =\frac{1}{\sqrt{\pi}}\sum_{n=1}^\infty
\frac{1}{n^{3/2}} [e^{i n\pi\kappa}-(-1)^n] e^{-n|\beta|} \, ,
\label{neg} \ee
i.e., for $\beta\rightarrow -\infty$:
\be C(\beta,\kappa) = (1/\sqrt{\pi}) e^{-|\beta|}(1+e^{i \pi
\kappa}). \ee
When the chemical potential and $\beta$ are positive, we divide the
integration range in (\ref{C}) so that for each interval one can use
the same expansions of the logarithms, and find
\[ C(\beta,\kappa)=\frac{2}{\pi}\Big[ i\pi(1-\kappa) \sqrt{\beta}+
\sum_{n=1}^\infty \frac{1}{n} \Big( [e^{-i n\pi\kappa}-(-1)^n]
\times \]

\vspace{-3ex}

\[ \int_0^{\sqrt{\beta}} e^{n(\lam^2-\beta)} d\lam + [e^{i
n\pi\kappa}-(-1)^n] \int_{\sqrt{\beta} }^\infty
e^{-n(\lam^2-\beta)}d\lam \Big)\Big] . \]
For large $\beta$, both integrals in this expression are evaluated
asymptotically as $1/(2n\sqrt{\beta})+O(\beta^{-3/2})$, and the
resulting sums over $n$ can be taken explicitly using the standard
formulae \cite{GR} $\sum_{k=1}^\infty(-1)^{n+1}/n^2=\pi^2/12$, and
$\sum_{k=1}^\infty\cos n\pi\kappa/n^2=\pi^2 B_2(\kappa/2)$, where
$B_2(x)=x^2-x+1/6$ is the second Bernoulli polynomial. This gives
$C(\beta,\kappa)$ for $\beta \rightarrow \infty$, i.e., in the
low-temperature limit $T\ll h$, as
\be C(\beta,\kappa) = \frac{\pi}{2\sqrt{\beta}}(1-\kappa)^2 +2i
\sqrt{\beta} (1-\kappa)\, . \label{low} \ee
The point $\beta=0$ is regular in Eq.~(\ref{C}), and the expansion for
small positive $\beta\rightarrow + 0$ coincides with the expansion
(\ref{neg}) for negative $\beta$, when $\beta\rightarrow -0$.


The regime of positive chemical potential and low temperatures,
$\beta\rightarrow\infty$, should correspond to the {\em conformal
limit}, when the asymptotics of the correlator reduces to the
bosonization result (\ref{bos2}). Using Eq.~(\ref{low}) for the
large-$\beta$ behavior of $C(\beta,\kappa)$ together with
Eq.~(\ref{lamp}) for $\lambda_j$ which gives in this limit:
\be  \lam_0= \sqrt{\beta} +i\frac{\pi\kappa}{2\sqrt{\beta}}\,
,\;\;\;\; \lam_{-1}= -\sqrt{\beta} +i \frac{\pi(2-\kappa) }{2
\sqrt{\beta}}\, , \label{int} \ee
we see that the exact asymptotic behavior  of the field-field anyonic
correlator (\ref{asympt}) reduces at low temperatures to the
following expression:
\begin{eqnarray}
\la \fad(x)\fa(0)\ra &\simeq& c_0 \, e^{-x(\pi T/v_F)
(\kappa^2+1)/2} e^{ixk_F\kappa} \label{conf} \\
& &+\, c_{-1} \, e^{-x(\pi T/v_F)[(\kappa-2)^2+1]/2}
e^{ik_F(\kappa-2)}. \nonumber
\end{eqnarray}
The asymptotics (\ref{bos2}) of the correlator within the
bosonization approach agrees at non-vanishing temperatures
$T\neq 0$ with the the exact expression (\ref{conf}). Indeed,
for $T\neq 0$, only one, the leading, exponent in
$\sinh(\pi Tx/v_F)$ dominates at
sufficiently large $x$, and the two expressions coincide. This means
that, taken together, the bosonization result (\ref{bos2}) and our
Eq.~(\ref{conf}) describe the correlator for all finite temperatures
starting with $T=0$, when Eq.~(\ref{bos2}) reduces to a combination
of powers of $x$.


Another important case reproduced by Eq.~(\ref{asympt}) is the
{\em bosonic limit} $\kappa \rightarrow 0$. As discussed above,
only the first term in (\ref{asympt}) needs to be taken into account
in this case. The asymptotics depend essentially on the sign of the
chemical potential $h$, i.e. $\beta$. For $\beta<0$, the argument of
the logarithm in Eq.~(\ref{C}) is positive and
\[ C(\beta,\kappa=0)= C(\beta) \equiv \frac{1}{\pi} \inti\ln \Big|
\frac{e^{\lam^2-\beta}+1}{e^{\lam^2-\beta}-1} \Big| d \lam \, . \]
For $\beta>0$, the argument of the logarithm is negative in the
interval $\lambda\in [-\sqrt{\beta}, \sqrt{\beta}]$, so that
\[ \lim_{\kappa\rightarrow 0} C(\beta,\kappa)=
C(\beta)+2i\sqrt{\beta}\, . \]
Also, at $\kappa= 0$ Eq.~(\ref{lamp}) gives
$\lam_0=\sqrt{\beta}$, and $\lam_0=i\sqrt{|\beta|}$ for positive and
negative $\beta$, respectively. Combining these equations, we see
that Eq.~(\ref{asympt}) reduces at $\kappa\rightarrow 0$ to
\be \la \fad(x)\fa(0)\ra \simeq c_0  \left\{ \begin{array}{ll} e^{-x
[\frac{\sqrt{T}}{2}C(h/T) +\sqrt{|h|}]}, & h<0\, , \\
e^{-x \frac{\sqrt{T}}{2}C(h/T)}, & h>0\, , \end{array} \right. \ee
and agrees with the results obtained previously for the impenetrable
bosons \cite{IIK2,IIK3}.


In the opposite {\em fermionic limit} $\kappa\rightarrow 1$, all
characteristics of the system (\ref{hama}), including the
correlators, should be those of the free fermions. Indeed,
Eqs.~(\ref{C}) and (\ref{lamp}) show that $C(\beta,\kappa)=0$ and
$\lambda_0 = -(\lambda_{-1})^*=(\beta+i\pi)^{1/2}$ for $\kappa= 1$,
and Eq.~(\ref{asympt}) reproduces the asymptotics of the field
correlator for free fermions. More generally, using the standard
Wigner-Jordan transformation, one can express the anyonic fields
(\ref{com1}) of arbitrary statistics $\kappa$ in terms of (in our
case free) fermions $\xi$ as
\[ \fa(x,t)=e^{i \pi (1-\kappa)n(x,t)}\xi(x,t), \;\;\; n(x,t) =\int^x
dx'\rho(x',t)  \, , \]
where $\rho=\xi^{\dagger}\xi$ is the particle density operator. This
representation provides a simple and useful interpretation of the
$C(\beta,\kappa)$ part of the asymptotics (\ref{asympt}). In terms
of the free fermions $\xi$, the anyonic correlator is:
\be \la \fad(x)\fa(0)\ra = \la \xi^{\dagger}(x) e^{i \pi
(\kappa-1)n_d } \xi(0) \ra , \label{fer2} \ee
where $n_d=n(x)-n(0)$ is the operator of the number of particles in
the interval $[0,\,x]$. This equation shows that the decay of the
correlator is caused in part by the fluctuations of the number of
particles $n_d$ creating the fluctuations of the accumulated
statistical phase of the correlator. The average taken over these
fluctuations leads to the correlator suppression. For large
distances $x$, one can treat the fluctuations of $n_d$ simply as
fluctuations of the number of free fermions in the system of size
$x$. They are governed by the Fermi distribution
$\vartheta(k)= 1/[e^{ (k^2 -h)/T}+1]$, which gives the probability
$\vartheta(k)$ for the particle with momentum $k$ to be present in
the system. Depending on whether each particle is in the interval
$[0,\,x]$ or not, its contribution to the phase factor $e^{i \pi
(\kappa-1)n_d}$ in the correlator (\ref{fer2}) is either
$e^{i \pi (\kappa-1)}$ or $1$. With this understanding,
\begin{eqnarray}  \la e^{i \pi (\kappa-1) n_d } \ra =\prod_{k}
[1-\vartheta(k) + e^{i \pi (\kappa-1)} \vartheta(k)]
\nonumber \\
= \exp \Big\{ \frac{x}{2\pi} \int d k \ln  \frac{ e^{(k^2 -h)/T} -
e^{i \pi \kappa}}{e^{(k^2 -h)/T}+1} \Big\} \, .  \label{fer3}
\end{eqnarray}
The change of variables from $k$ to $\lam$ shows that this is
exactly the $C(\beta,\kappa)$ part of the asymptotics (\ref{asympt}).


Finally, we present an {\em outline of the calculations} leading to
the asymptotics (\ref{asympt}) of the anyonic field-field
correlator. The starting point of these calculations is the
representation of the correlator as the Fredholm determinant   of an
integral operator which we have obtained previously by two different
methods: the anyonic generalization \cite{SSC,PKA2} of Lenard's formula
\cite{L1}, and direct summation of the form-factors \cite{PKA3}. The
kernel of the integral operator of interest can be written as
\be\label{kern} K_T(\lam,\mu)= \frac{e_+(\lam)e_-(\mu)-e_- (\lam)e_+
(\mu)}{2i (\lam-\mu)} \, , \ee
where $e_{\pm} (\lam) =\sqrt{\vartheta(\lam)}e^{\pm i\lam x}$. Here,
and in the remainder of the paper, to simplify the notations we use
the coordinates rescaled by the temperature-dependent factor: $x
\equiv (x_1-x_2) \sqrt{T} /2$.

The integral operators with the kernels of the form (\ref{kern})
represent a special type of the general class of ``integrable"
operators, which characterize a large set of correlation functions
of integrable models and spectral distributions for random matrix
ensembles \cite{WMTB,JMMS,IIKS,IIKV,KBI,HI}. An important property
of the integrable operators is that the resolvent of the operator is
also integrable. In the case of the kernel (\ref{kern}), the
resolvent kernel can be can be written in the same form through the
two functions $f_\pm (\lam)$ introduced as solutions of the integral
equations
\be f_\pm(\lam)-\gamma \int_{-\infty}^{+\infty} K_T(\lam,\mu)
f_\pm(\mu) d \mu=e_\pm(\lam)\, , \ee
where $\gamma \equiv (1+e^{i\pi\kappa})/\pi$. In terms of these
functions, one can define the auxiliary ``potentials'' $B_{l,m}$ as
\be B_{lm}(x,\beta,\kappa) \equiv \gamma \int_{-\infty}^{+\infty}
e_l(\lam) f_m(\lam)\ d\lam\, , \ \ l,m=\pm\, . \ee
With these notations, the result for the determinant representation
of the field correlator derived in \cite{PKA2,PKA3} is expressed
as
\be \label{templ} \la \fad(x_1)\fa(x_2)\ra
=B_{++}(x,\beta,\kappa)e^{\sigma(x,\beta,\kappa)}\, , \ee
where $\sigma=\ln \det(1-\gamma \hat K_T)$ is the logarithm of the
Fredholm determinant.

Following the method developed for impenetrable bosons
\cite{IIK2,IIK3,KBI}, we have obtained the integrable system of partial
nonlinear differential equations for the potentials $B_{++}$ and
$B_{+-}$:
\be \label{ints}
\partial_\beta B_{+-} =x+\frac{1}{2}\frac{\partial_x \partial_
\beta B_{++}}{B_{++}}\, ,\;\;\; \partial_x B_{+-} = B_{++}^2\, , \ee
with the initial conditions at fixed $\beta$ and $x\rightarrow 0$:
\be \label{incondi} B_{++}(x,\beta,\kappa) = B_{+-}(x,\beta,\kappa)=
\gamma d(\beta)+[\gamma d(\beta)]^2x\, , \ee
where $d(\beta)=\inti\vartheta(\lam)d\lam$, and the condition
\begin{eqnarray*}
B_{++}(x,\beta,\kappa)=B_{+-}(x,\beta,\kappa)=0\, ,\;\;\; \mbox{for}
\; \beta \rightarrow -\infty\, .
\end{eqnarray*}
The derivatives of the logarithm of the determinant $\sigma$ are
\begin{eqnarray}\label{desigma}
\partial_x\sigma&=&-B_{+-}\, , \ \ \ \partial_x^2\sigma=-B_{++}^2\, ,
\nonumber \\
\partial_\beta \sigma &=&-x\partial_\beta B_{+-}+\frac{1}{2}(\partial_
\beta B_{+-})^2 -\frac{1}{2}(\partial_\beta B_{++})^2 ,
\end{eqnarray}
which shows that the field-field correlator can be expressed in
terms of the solutions of the integrable system (\ref{ints}). Also,
$\sigma$ satisfies the equation
\be (\partial_\beta\partial_x^2
\sigma)^2+4(\partial_x^2\sigma)[2x\partial_\beta \partial_x\sigma+
(\partial_\beta\partial_x\sigma)^2-2\partial_\beta\sigma]=0 \ee
with initial conditions
\begin{eqnarray*}
\sigma&=&-\gamma d(\beta)x-[\gamma d(\beta)]^2\frac{x^2}{2}+O(x^3)\,
, \ \ x\rightarrow 0\, \\ \sigma&=&0\, ,\ \  \beta\rightarrow
-\infty\, .
\end{eqnarray*}
The integrable system (\ref{ints}) is the same system that
characterizes impenetrable bosons, but with different initial
conditions (\ref{incondi}). At $\kappa=0$, i.e. $\gamma=2/\pi$, it
reproduces the bosonic result \cite{IIK2,IIK3,KBI}.
The same phenomenon was observed by Santachiara and
Calabrese \cite{SC1} (see also \cite{SC2}) who showed that the field correlator at zero temperature in the finite box
satisfies the same Painlev\'e VI differential equation characterizing impenetrable bosons but with statistics
dependent initial conditions.

As the final step, the large distance asymptotics of the potentials
$B_{++},B_{+-}$ were found from the matrix Riemann-Hilbert problem
associated with the integrable system (\ref{ints}). The technique
is similar to the one developed for impenetrable bosons
\cite{IIK2,IIK3,KBI}. The asymptotic behavior of the potentials,
combined with Eqs.~(\ref{templ}) and (\ref{desigma}), gives the main
result of this work, the large distance asymptotics (\ref{asympt})
of the field-field correlator for impenetrable one-dimensional anyons.

V.E.K. would like to thank P. Calabrese for useful discussions. The
authors were supported by the NSF grants \# DMR-0325551, PHY-0653342,
and DMS-0503712.


\end{document}